\documentclass[flEq,usenatbib]{mnras}
\usepackage{newtxtext,newtxmath}
\usepackage[T1]{fontenc}
\DeclareRobustCommand{\VAN}[3]{#2}
\let\VANthebibliography\thebibliography
\def\thebibliography{\DeclareRobustCommand{\VAN}[3]{##3}\VANthebibliography}
\usepackage{graphicx}	
\usepackage{amsmath}	
\usepackage{bm}
\renewcommand{\Vec}{\bm}
\newcommand{\changed}[1]{#1}

\title[\textit{LISA} sensitivity to EMRIs]{Rapid determination of \textit{LISA} sensitivity to extreme mass ratio inspirals with machine learning}

\author[C. E. A. Chapman-Bird et al.]{
Christian E. A. Chapman-Bird,$^{1}$\thanks{E-mail: c.chapman-bird.1@research.gla.ac.uk}
Christopher P. L. Berry$^{1}$ and Graham Woan$^{1}$
\\
$^{1}$SUPA, University of Glasgow, Glasgow G12 8QQ, UK\\
}

\date{Accepted 2023 May 5. Received 2023 May 4; in original form 2023 January 26}

\pubyear{202\changed{3}}

\begin{document}
\label{firstpage}
\pagerange{\pageref{firstpage}--\pageref{lastpage}}
\maketitle

\begin{abstract}
Gravitational wave observations of the inspiral of stellar-mass compact objects into massive black holes (MBHs), extreme mass ratio inspirals (EMRIs), enable precision measurements of parameters such as the MBH mass and spin.
The \textit{Laser Interferometer Space Antenna} is expected to detect sufficient EMRIs to probe the underlying source population, testing theories of the formation and evolution of MBHs and their environments.
Population studies are subject to selection effects that vary across the EMRI parameter space, which bias inference results if unaccounted for.
This bias can be corrected, but evaluating the detectability of many EMRI signals is computationally expensive. 
We mitigate this cost by (i) constructing a rapid and accurate neural network interpolator capable of predicting the signal-to-noise ratio of an EMRI from its parameters, and (ii) further accelerating detectability estimation with a neural network that learns the selection function, leveraging our first neural network for data generation.
The resulting framework rapidly estimates the selection function, enabling a full treatment of EMRI detectability in population inference analyses. 
We apply our method to an astrophysically\changed{-}motivated EMRI population model, demonstrating the potential selection biases and subsequently correcting for them.
Accounting for selection effects, we predict that \changed{with 116 EMRI detections} \textit{LISA} will measure the MBH mass function slope to a precision of $8.8\%$, the CO mass function slope to a precision of $4.6\%$, the width of the MBH spin magnitude distribution to a precision of $10\%$ and the event rate to a precision of $12\%$ with EMRIs at redshifts below $z=6$.

\end{abstract}


\section{Introduction}

The \textit{Laser Interferometer Space Antenna} \citep[\textit{LISA};][]{Amaro-Seoane2017} is a planned space-based observatory capable of observing gravitational waves (GWs) with frequencies $\sim10^{-5}$--$10^{-1}~\mathrm{Hz}$. 
A promising source of GWs in this band are extreme mass ratio inspiral (EMRI) systems, comprising a compact object (CO) orbiting, and gradually inspiraling towards, a massive black hole \citep[MBH;][]{Amaro-Seoane2007,Amaro-Seoane2022}. 
EMRI systems typically complete $\sim10^4\text{--}10^5$ orbital cycles in the \textit{LISA} band and generate GWs with an intricate frequency evolution. 
Relativistic effects, including Lense--Thirring precession and pericentre precession, generate many distinct sideband modes in the signal spectrum \citep{Hughes2021, Barack2004}. 
The amplitude and phase evolution of these modes is strongly dependent on the parameters of the MBH--CO system, enabling precise measurements of these parameters \citep{Babak2017}. 
In particular, the (redshifted) MBH and CO masses, MBH spin and orbital eccentricity may be determined to accuracies of $\sim10^{-3}\%$, and \changed{the sources} localised in space to better than $10\%$ relative precision \citep{Amaro-Seoane2007,Berry2019}.
This level of measurement precision for MBHs surpasses both existing electromagnetic techniques \citep{Daly2011} and \textit{LISA} observations of MBH binaries \citep{Klein2016}.

The number of EMRIs that will be detected is uncertain, largely due to poorly constrained astrophysical parameters in current formation channel theories, but the detection rate is likely to be of order $10^1\text{--}10^3~\mathrm{yr^{-1}}$ \citep{Babak2017,Pan2021,Vazquez-Aceves2021,Amaro-Seoane2022}. 
During \textit{LISA}'s proposed $4~\mathrm{yr}$ mission, we therefore expect to attain a sufficiently large catalogue of EMRIs (each of which providing excellent measurements of their parameters) to probe and resolve features of the MBH mass and spin populations. 
For example, we expect to match the precision of current observational estimates of an MBH mass power-law spectral index, even for pessimistic EMRI detection rate predictions \citep{Gair2010}.
Such a catalogue also enables the testing of the wide array of EMRI formation channel theories.
Several EMRI formation channels have been suggested, including loss-cone scattering of COs into inspiral orbits \citep{Alexander2017}, radial migration of COs by dynamical interaction with an accretion disc \citep{Pan2021}, capture via the Kozai--Lidov mechanism due to the presence of a binary MBH system \citep{Naoz2022}, and the tidal disruption of main-sequence or helium stars on highly eccentric orbital trajectories around MBHs \citep{Bortolas2019}.
For a given formation channel, the rate of EMRI production depends on astrophysical parameters \citep[e.g.,][for loss-cone or TDE channels]{Broggi2022} that are poorly constrained by existing observations.
Determining the relative contributions of the formation channels to the overall EMRI rate therefore places constraints on these parameters; this could be estimated from a catalogue of EMRI observations in a similar manner to how ground-based GW observations are used to constrain their source astrophysics \citep[e.g.,][]{Zevin2021}.

Extracting information about the astrophysical population requires a hierarchical inference where the parameters of each EMRI in the catalogue (and their associated uncertainties) are collectively used to constrain the parameters of a chosen population model. 
However, the catalogue only contains sources loud enough to cross a detection threshold, and these are generally not representative of the underlying population: if these selection effects are ignored, the inferred population parameters will be biased.
EMRIs that are intrinsically fainter (depending on their masses, spins or orbital parameters) and more distant are less likely to be detected than their nearer and louder counterparts. 
In practice, one may correct for this by determining the fraction of the proposed population that is detectable, and re-weighting the population likelihood accordingly \citep{Mandel2018, Alsing2022}. 

In the absence of a standard EMRI detection pipeline, the detectability of a given EMRI is typically assessed with respect to the signal-to-noise ratio (SNR) of the EMRI waveform \citep{Gair2004,Gair2010,Babak2017}. 
The detectable fraction of a proposed population (the selection function) may therefore be estimated by randomly drawing EMRI events from the population and computing their SNRs to determine the fraction of these samples that are detectable. 
These selection function estimates may then be used to re-weight the population likelihood and obtain unbiased inferences of the population parameters.

The function that maps EMRI parameters to the waveform SNR is complicated, so bias correction is computationally expensive. 
This high cost comes from both the generation of complex long-duration waveforms and the manipulation of these large data sets.  
Even exploiting graphics processing unit (GPU) acceleration and vectorisation, each SNR takes of order $1~\mathrm{s}$ to compute for a $4~\mathrm{yr}$ data-stream duration.
Using conservative estimates, if one draws $10^5$ EMRIs for each candidate population, and a population inference sampling run consists of $10^5$ candidate populations, full selection bias correction would require $10^{10}$ GPU$~\mathrm{s}$.
This is too costly for analyses including selection bias correction in this manner to be conducted in a reasonable amount of time.

Previous studies have addressed the issue of computational cost by indirectly approximating the behaviour of the selection function via\changed{:} proxy distance thresholds \citep{Laghi2021}\changed{;} a reduction in EMRI parameter space complexity, for instance by neglecting the dependence of eccentricity or inclination on detectability \citep{Gair2010}; \changed{a reduction of waveform complexity by only computing a small number of sideband modes with faster, less accurate waveform models \citep{Chua2022}}.
These approaches permit the rapid computation of the selection function, but do not account for more complex correlations between EMRI parameters and may introduce systematic biases due to the approximations made. 
For instance, the evolution of orbital eccentricity and inclination during an inspiral is correlated with mass ratio, and the mode amplitudes (and therefore, the overall SNR of the waveform) are correlated with both eccentricity and inclination evolution. 
Therefore, even for parameters not directly of interest to a given population study, the correlations between these parameters and event detectability must still be taken into account to avoid biases in the results obtained.

In this work, we propose an alternative approach that leverages the speed of the recently developed GPU-accelerated EMRI waveform package FastEMRIWaveforms \citep[FEW;][]{Katz2021} and machine-learning techniques to interpolate the EMRI SNR function, directly correcting for an SNR-based selection bias without the need for major systematic approximations or simplifications.
In Section~\ref{sec:framework}, we outline the Bayesian population inference framework employed, including the correction for selection biases. 
In Section~\ref{sec:snr_interp}, we introduce our approach for estimating the selection bias with machine learning, which we achieve by replacing the bottleneck in the selection bias calculation (the EMRI SNR function) with a neural network interpolator.
The effectiveness of our framework is demonstrated in Section~\ref{sec:sampling} for a typical EMRI population, presenting clearly the manifestation of the selection bias in the obtained results and how this is corrected for in practice; the corrected results provide an unbiased estimate for the precision to which \textit{LISA} observations could constrain the astrophysical EMRI population.
Finally, in Section~\ref{sec:pp}, we perform a global posterior consistency check to verify the analysis.

The method we describe in this paper is implemented in our open-source code package \texttt{poplar} \citep{poplar}, which we plan to use for (and develop alongside) future EMRI population studies.

\section{Hierarchical Bayesian inference framework}
\label{sec:framework}

Our goal is to infer the properties of an EMRI population model using a catalogue of many EMRI observations.

For each EMRI in the catalogue, information about its parameters $\Vec{\theta}_i$ is encoded in the data, $\Vec{d}$, where the subscript $i \in [1, N_\mathrm{obs}]$ identifies the particular EMRI in the catalogue of $N_\mathrm{obs}$ detections. 
The posterior distribution for $\Vec{\theta}_i$ given the data is
\begin{equation}
    p(\Vec{\theta}_i | \Vec{d}) = \frac{\pi(\Vec{\theta}_i)\mathcal{L}(\Vec{d}|\Vec{\theta}_i)}{\mathcal{Z}(\Vec{d})},
\end{equation}
where $\pi(\Vec{\theta}_i)$ is the prior distribution on $\Vec{\theta}_i$, $\mathcal{L}(\Vec{d}|\Vec{\theta}_i)$ is the likelihood of observing the data given a set of source parameters, and $\mathcal{Z}(\Vec{d})$ is the evidence (marginalised likelihood).
We estimate the parameters of the EMRI by stochastically sampling the posterior distribution, obtaining a set of posterior samples $\{^k\theta_i\}$ \citep{Christensen2022}, where the superscript $k \in [1, S_i]$ denotes each posterior sample for a given event. 
The posterior $p(\Vec{\theta}_i | \Vec{d})$ provides information about a single EMRI source; by combining together the properties of the catalogue of sources,we can constrain a population model.

The population model $p_\mathrm{pop}(\Vec{\theta}|\Vec{\lambda})$ describes the astrophysical distribution of EMRI source parameters. 
It is described by a set of hyperparameters $\Vec{\lambda}$ that determine the shape of the population, and a Poissonian mean event rate $\mathcal{R}$ that parameterises how often EMRIs occur.
We use the \emph{hyper} prefix to differentiate these population-level (hyper)parameters from the event-level EMRI parameters. 
By estimating the hyperparameters, we constrain the relative probabilities of different population shapes and event rates in accordance with the contents of the catalogue. 
We perform this hyperparameter estimation in a hierarchical Bayesian inference framework \citep{Mandel2018}.

To obtain an estimate of $\Vec{\lambda}$ and $\mathcal{R}$, we sample the hyperparameter posterior distribution
\begin{equation}
    p(\Vec{\lambda}, \mathcal{R} | \{\Vec{\theta}\}) = \frac{\pi(\Vec{\lambda}) \pi(\mathcal{R}) \mathcal{L}(\{\Vec{\theta}\}|\Vec{\lambda}) \mathcal{L}(\{\Vec{\theta}\}|\mathcal{R})}{\mathcal{Z}(\{\Vec{\theta}\})},
    \label{eq:hyperposterior}
\end{equation}
where $\pi(\Vec{\lambda})$ and $\pi(\mathcal{R})$ are hyperprior distributions, and $\mathcal{Z}(\{\Vec{\theta}\})$ is the hyperevidence. 
The hyperparameter likelihood $\mathcal{L}(\{\Vec{\theta}\}|\Vec{\lambda})$ is defined as
\begin{equation}
    \mathcal{L}(\{\Vec{\theta}\}|\Vec{\lambda}) = \prod_{i=1}^{N_\mathrm{obs}} \frac{1}{S_i \alpha(\Vec{\lambda})}\sum_{j=1}^{S_i}\frac{p_\mathrm{pop}(^j\Vec{\theta}_i|\Vec{\lambda})}{\pi(\Vec{\theta}_i)},
    \label{eq:hp_likelihood}
\end{equation}
where. in general, the population probability of each posterior sample must be re-weighted by the prior used in the EMRI parameter estimation step \citep{Mandel2018}.
In our case, we adopt uniform priors on all EMRI parameters and this re-weighting simplifies to a proportionality constant. 
The rate likelihood $\mathcal{L}(\{\Vec{\theta}\}|\mathcal{R})$ is 
\begin{equation}
    \mathcal{L}(\{\Vec{\theta}\}|\mathcal{R}) = \exp[{-\mathcal{R}\alpha(\Vec{\lambda})}]\left[\mathcal{R}\alpha(\Vec{\lambda})\right]^{N_\mathrm{obs}}.
    \label{eq:rate_likelihood}
\end{equation}
Here, the selection function $\alpha(\Vec{\lambda})$ is a corrective factor applied to account for the presence of selection bias on the observations: of the (unknown) number of events that occurred, only a subset $N_\mathrm{obs}$ were detected. 
It may be written as 
\begin{equation}
    \alpha(\Vec{\lambda}) = \int p_\mathrm{det}(\Vec{\theta}) p_\mathrm{pop}(\Vec{\theta} | \Vec{\lambda})\,\mathrm{d}\theta,
    \label{eq:selfun}
\end{equation} 
for some detection probability $p_\mathrm{det}(\Vec{\theta})$, and represents the fraction of a population (described by a particular set of hyperparameters $\Vec{\lambda}$) that is detectable. 
Performing an inference including the selection effects should produce results unbiased by detectability \citep{Mandel2018}.

To use this inference framework, one must specify:
\begin{enumerate}
  \item A method for obtaining $\{\Vec{\theta}\}$;
  \item The form of $p_\mathrm{pop}(\Vec{\theta} | \Vec{\lambda})$;
  \item The selection effects for the observations.
\end{enumerate}
Once these three ingredients have been formally defined, we can construct (and ultimately, sample) the hyperposterior, Eq.~\eqref{eq:hyperposterior}.
We introduce the approach we use to obtain posterior samples for each detected EMRI in Section~\ref{sec:emriPE}; we
detail the form of our population model in Section~\ref{sec:pop}, and we outline our treatment of detectability and the modelling of selection effects in Section~\ref{sec:selection_effects}.

\subsection{EMRI parameter estimation}
\label{sec:emriPE}

Despite recent reductions in EMRI waveform computation time to the sub-second level \citep{Katz2021}, standard Bayesian parameter estimation techniques are too costly for the event posteriors to be sampled directly en-masse as is required in population studies. 
We instead opt to approximate the EMRI likelihood (and by extension, posterior) with a Fisher matrix approach, operating under the linear signal approximation \citep[LSA;][]{Cutler1994}, in which the likelihood is approximated by a multivariate normal distribution,  
\begin{equation}
    p(d | \Vec{\theta}) \approx \mathcal{N}\left(\Vec{\theta},\Gamma^{-1}\right),
\end{equation}
where the covariance matrix of the distribution is the inverse of the Fisher information matrix (FIM) $\Gamma$ of the EMRI waveform.
The LSA is only valid in the high-SNR limit (which may lie far above the detection threshold), which should be verified before it is used to approximate likelihoods \citep{Vallisneri2008}. 
The FIM is given by
\begin{equation}
    \Gamma_{\ell m} = \langle \partial_\ell h | \partial_m h \rangle, 
\end{equation}
where $\partial_\ell h$ refers to the derivative of the waveform strain $h(t)$ with respect to the $\ell$-th parameter of $\Vec{\theta}$, evaluated at $\Vec{\theta}$.
The noise-weighted inner product is defined as
\begin{equation}
     \langle x | y \rangle = 4 \mathfrak{R} \left[\int_0^\infty \frac{\Tilde{x}^*(f)\Tilde{y}(f)}{S_n(f)}\mathrm{d}f\right],
\end{equation}
where $\Tilde{x}(f)$ is the Fourier transform of a time-domain strain $x(t)$, $\mathfrak{R}$ refers to the real part and $S_n(f)$ is the one-sided power spectral density (PSD) of the detector \citep{Maggiore}.
We adopt the analytic fit to the \textit{LISA} PSD derived in \citet{Robson2019}.
Using the FIM, we can rapidly produce posterior distributions for a catalogue of EMRIs.

Specifics of our EMRI event catalogue simulation pipeline, including both waveform generation and FIM computation, are discussed in Appendix~\ref{sec:IndivEMRIs}.

\subsection{Population model}
\label{sec:pop}

For simplicity, we choose a population model that is a product of independent univariate subpopulations, such that
\begin{equation}
    p_\mathrm{pop}(\bm{\theta} | \bm{\lambda}) = \prod_{x \in \bm{\theta}} p_x (x | \Vec{\lambda}_x),
\end{equation}
where $x$ denote EMRI parameters and $\Vec{\lambda}_x$ the corresponding hyperparameters that describe the shape of the subpopulation.
The mathematical form of these subpopulations is summarised in Table~\ref{tab:populations}, and in detail:
\begin{itemize}
    \item 
    Mass functions for both MBHs and stellar-mass black holes are well approximated by power laws, albeit with additional substructure present when examined in detail \citep{Shankar2013, o3pop}.
    We therefore model the MBH and CO mass distributions $p_M(M | \lambda_M, M_\mathrm{min}, M_\mathrm{max})$ and $p_\mu(\mu | \lambda_\mu, \mu_\mathrm{min}, \mu_\mathrm{max})$ as power laws, with index $\lambda_x$ and limits $[x_\mathrm{min},x_\mathrm{max}]$, which have the form
    \begin{equation}
        p_x(x | \lambda_x, x_\mathrm{min}, x_\mathrm{max}) = \frac{1 + \lambda_x}{x_\mathrm{max}^{1+\lambda_x} - x_\mathrm{min}^{1+\lambda_x}} x^{\lambda_x}.
    \end{equation}
    
    \item
    The form of the MBH spin magnitude distribution $p_a(a | \mu_a, \sigma_a)$ is dependent on a number of astrophysical processes during the formation and evolution of MBHs and their host galaxies \citep{Volonteri2010, Sesana2014}.
    Incorporating these into our population model and characterising their impact on inference results is beyond the scope of this study.
    For simplicity, we instead choose a truncated normal distribution with mean $\mu_a$ and variance $\sigma_a^2$ as has been done in previous analyses of the stellar-mass binary black hole (BBH) mergers \citep{Roulet2019,Miller2020}.
    This is written as
    \begin{equation}
        p_x(x | \mu_x, \sigma_x) = \frac{1}{\sigma}\frac{\psi\left[(x - \mu_x)/{\sigma_x}\right]}{\Psi\left[({B - \mu_x})/{\sigma_x}\right] - \Psi\left[({A - \mu_x})/{\sigma_x}\right]}, 
    \end{equation}
    where $\psi(x)$ and $\Psi(x)$ are the probability density and cumulative distribution functions (CDF) of the standard normal distribution, respectively. 
    The limits $[A,B]$ are chosen to be $[0.001, 0.999]$ as waveform generation is unstable at extremal spins beyond these limits.
    \item
    High initial orbital eccentricities ($>0.99$) are expected for EMRIs formed by relaxation mechanisms, but significant orbital eccentricity will be lost before the GW emission of the system enters the \textit{LISA} band \citep{Peters1963}, broadening the distribution and shifting it to lower eccentricities \citep{Amaro-Seoane2020}.
    To reflect this behaviour, we choose a uniform eccentricity distribution $e_0 \in [0.1, 0.5]$, with upper limit chosen to reflect that the waveform model is a series expansion in eccentricity and should therefore not be trusted for high eccentricities \citep{Fujita2020, Isoyama2022}.
    The waveform model also consists of a system of ordinary differential equations (ODEs) that must be solved \citep{Katz2021}. 
    The lower limit of the eccentricity distribution is chosen due to increasing stiffness in this ODE system at lower eccentricities leading to high computational cost \citep{Burden1993}. 
    We do not anticipate the validity of our approach to be affected by this lower eccentricity cut-off. 
    \item
    Orbital inclination $\iota_0$ is similarly truncated due to ODE stiffness issues, but is otherwise distributed uniformly on the unit sphere along with other angular parameters. 
    \item
    We choose a redshift distribution that is uniform in comoving volume and in comoving time \citep{Hogg1999}; this has the form
    \begin{equation}
        p_z(z) \propto \frac{1}{(1+z)E(z)}\left(\int_0^{z} \frac{\mathrm{d}z'}{E(z')}\right)^2,
    \end{equation}
    where
    \begin{equation}
        E(z) = \sqrt{\Omega_M (1+z)^3  + \Omega_\Lambda},
    \end{equation}
    and we assume a standard cosmology with $\Omega_M = 0.3$ and $\Omega_\Lambda = 1 - \Omega_M = 0.7$. 
    The upper redshift limit for this distribution is chosen to be $z=6$ such that the detectable region of parameter space is not significantly truncated (otherwise, selection effects will be artificially suppressed).
    Increasing the redshift limit leads to high computational costs as the event rate $\mathcal{R}$ must also be increased accordingly, as the event rate density has remained constant but the comoving volume over which we are distributing events has grown.
\end{itemize}
\begin{table}
	\centering
	\caption{The functions, free parameters, and limits of the sub-population distributions $p_x(x | \Vec{\lambda}_x)$, the product of which is the EMRI population chosen. 
	The hyperparameters $\Vec{\lambda}_x$ are estimated via population inference. 
	The upper limit for $t_\mathrm{plunge}$ is reduced to $2~\mathrm{yr}$ for our validation analysis (Section~\ref{sec:pp}).} 
	\label{tab:populations}
	\begin{tabular}{llll} 
		\hline
		$x$ & $p_x(x | \Vec{\lambda}_x)$ & $\Vec{\lambda}_x$ & $[x_\mathrm{min}, x_\mathrm{max}]$\\
		\hline
		$M$ & Power-law & $\lambda_M$, $M_\mathrm{min}$, $M_\mathrm{max}$  & [$M_\mathrm{min}$, $M_\mathrm{max}$]\\
		$\mu$ & Power-law & $\lambda_\mu$, $\mu_\mathrm{min}$, $\mu_\mathrm{max}$ & [$\mu_\mathrm{min}$, $\mu_\mathrm{max}$]\\
		$a$ & Trunc. Normal & $\mu_a$,$\sigma_a^2$ & $[0.001, 0.999]$\\
		$e_0$ & Uniform & --- & [0.1, 0.5]\\		
		$\cos\iota_0$ & Uniform & --- & $[0, \pi/3]$\\		
		$\sin\theta_S$ & Uniform & --- & $[0, \pi]$\\		
		$\sin\theta_K$ & Uniform & --- & $[0, \pi]$\\		
		$\Delta\phi$ & Uniform & --- & $[0, 2\pi]$\\		
		$t_\mathrm{plunge}$ & Uniform & --- & $[0, (2,10)]$\,yr\\
		$z$ & $p_z(z)$ & --- & $[0, 6]$\\ %
		\hline
	\end{tabular}
\end{table}
The chosen form of these subpopulations is motivated primarily by computational simplicity.
However, our approach is flexible and can be applied to the hierarchical inference of any population model.

\subsection{Selection effects} \label{sec:selection_effects}

In the absence of a specific EMRI search pipeline, we model the detection probability as a binary SNR threshold, as is typical for EMRI studies \citep{Gair2010, Babak2017,Bonetti2020}. 
This may be written as
\begin{equation}
    p_\mathrm{det}(\Vec{\theta}) = \mathcal{H}(\rho_\mathrm{n} - \rho_\mathrm{t}),
    \label{eq:alpha_eq1}
\end{equation}
where $\rho_\mathrm{n}$ is a (noise-realized) SNR, $\mathcal{H}(x)$ is the Heaviside step function and $\rho_\mathrm{t}$ is a chosen threshold SNR. 
We obtain $\rho_\mathrm{n}^2$ by drawing a sample from a non-central $\chi^2$ distribution with two degrees of freedom and non-centrality parameter $\rho_\mathrm{opt}^2$ \citep{Maggiore},
\begin{equation}
    p\left(\rho_\mathrm{n}^\changed{2} \middle| \rho_\mathrm{opt}^\changed{2} \right) = \frac{1}{2}\exp\left(-\frac{\rho_\mathrm{n}^2 + \rho_\mathrm{opt}^2}{2}\right)I_0\left(\rho_\mathrm{n}\rho_\mathrm{opt}\right),
    \label{eq:simp_chi2}
\end{equation}
where $I_0(x)$ is a modified Bessel function of the first kind \citep{Abramowitz1964}, and $\rho_\mathrm{opt}^2$ is the square of the optimal matched-filter SNR
\begin{equation}
    \rho_\mathrm{opt}^2 = \langle h | h \rangle.
    \label{eq:opt_snr}
\end{equation}
\changed{We assume that $\rho_\mathrm{n}$ is the positive square root of $\rho_\mathrm{n}^2$; while noise fluctuations can lead to negative values, this is not expected for large values around our detection threshold.} 
One may analytically compute the mean detection probability in Eq.~\eqref{eq:alpha_eq1} over all noise realizations by directly computing the non-central $\chi^2$ CDF \changed{$p(\rho_\mathrm{n}^2 > \rho_\mathrm{t}^2 | \rho_\mathrm{opt}^2)$}, such that
\begin{equation}
    \overline{p_\mathrm{det}}(\Vec{\theta}) = 1 - p\left(\rho_\mathrm{n}^2 > \rho_\mathrm{t}^2 \middle| \rho_\mathrm{opt}^2\right),
    \label{eq:pdet_avg}
\end{equation}
where an overline denotes the mean.

We approximate the selection function Eq.~\eqref{eq:selfun} by evaluating the Monte Carlo sum
\begin{equation}
    \alpha(\Vec{\lambda}) \approx \frac{1}{N_\mathrm{t}}\sum_{k=0}^{N_\mathrm{t}} \overline{p_\mathrm{det}}(\Vec{\theta}_k),
    \label{eq:alpha_eq2}
\end{equation}
where $\{\bm{\theta}_k\}$ are sampled from $p_\mathrm{pop}(\bm{\theta}|\bm{\lambda})$. 
As the variance on this approximation scales inversely with $N_\mathrm{t}$, one must compute $\overline{p_\mathrm{det}}(\Vec{\theta})$ (and therefore $\rho_\mathrm{opt}$) of the order of $10^5$ times for each computation of $\alpha(\bm{\lambda})$ to achieve percent-level accuracy; even with parallelisation, this would be prohibitively expensive with typical computing resources (taking of the order of minutes) for use in a typical sampling run, in which $\alpha(\bm{\lambda})$ must be computed once per hyperlikelihood call. 
We address this problem by replacing the SNR function with an accurate and rapid interpolator, allowing for Eq.~\eqref{eq:alpha_eq2} to be evaluated in parallel at a sufficiently low computational cost to be practical for use in inference problems.

\section{Interpolating over signal-to-noise ratio}
\label{sec:snr_interp}

The principal requirements for our SNR interpolator are that it must be accurate and unbiased across the EMRI parameter space: inaccuracies may bias the results of our population inference.
It must also be sufficiently fast as to not bottleneck the sampling process, capable of estimating SNRs for $10^5$ sets of EMRI parameters in $<1~\mathrm{s}$. 
These constraints are particularly challenging to meet due to the high dimensionality of the EMRI parameter space (13 dimensions, as defined in Appendix~\ref{sec:IndivEMRIs}). 

Fortunately, we can reduce the number of parameters that we need to interpolate over by considering how the SNR of an EMRI waveform depends on each parameter. 
We can ignore the orbital phase parameters $(\Phi_r, \Phi_\theta,\Phi_\phi)$ due to their negligible correlation with SNR, as the initial phase becomes relatively unimportant for an inspiral with $\sim10^4$ orbital cycles. 
Additionally, as SNR scales inversely with luminosity distance $d_\mathrm{L}$, we may further reduce the dimensionality of the parameter space by fixing $d_\mathrm{L}$ in training data and applying this scaling post-interpolation: we use $d_\mathrm{L} = 1~\mathrm{Gpc}$ for convenience.

Despite eliminating four dimensions of the EMRI parameter space, we are still in a regime where standard interpolation schemes are ineffective.
As a representative example, we consider spline interpolation schemes with piecewise polynomials of zeroth, first, and third order: these are more commonly known as nearest neighbour, linear, and cubic spline interpolation respectively \citep{Piegl1987}.
Our requirement for a fast interpolator prevents us from interpolating over points randomly distributed in the parameter space, as the complexity of the algorithms used for this scales quadratically with the number of basis points \citep{Barber2013} and the computational cost of these methods quickly becomes impractical.
Instead, we may use grid-based versions of these techniques.
However, these methods suffer from the curse of dimensionality: the Euclidean distances between neighbouring grid verticies grows as the dimensionality of the space increases, which leads to poor interpolation accuracy.

To demonstrate the unsuitability of linear interpolation in practice, we generate $\rho_\mathrm{opt}$ on a regular grid with $10^6$ total grid points and construct the aforementioned spline interpolators with this grid as a basis.
We then compute a testing set of $10^6$ SNRs from randomly sampled sets of EMRI parameters and compare the interpolator output at these points by calculating the discrepancies between the true and predicted SNRs, denoted $\rho_\mathrm{true}$ and $\rho_\mathrm{pred}$, respectively. 
The cumulative distribution of the (i) absolute and (ii) relative differences between prediction and truth for the three spline interpolation schemes we consider are shown in Figure~\ref{fig:interps}. 
As expected, the grid-based interpolator performance is poor regardless of the order of the piecewise polynomial used. 
While linear or spline interpolation offers marginal improvement over nearest neighbour interpolation, the low spatial resolution of the grid limits the improvements.
The majority of the interpolated SNRs are inaccurate by at least $50\%$, with absolute errors typically exceeding $10$ (or even as high as $100$ in extreme cases). 
As we will demonstrate in Section~\ref{sec:pp}, this performance is inadequate for unbiased population inference.

\subsection{Interpolation with neural networks}
\label{sec:NNs}

\begin{figure*}
	\includegraphics[width=0.7\paperwidth]{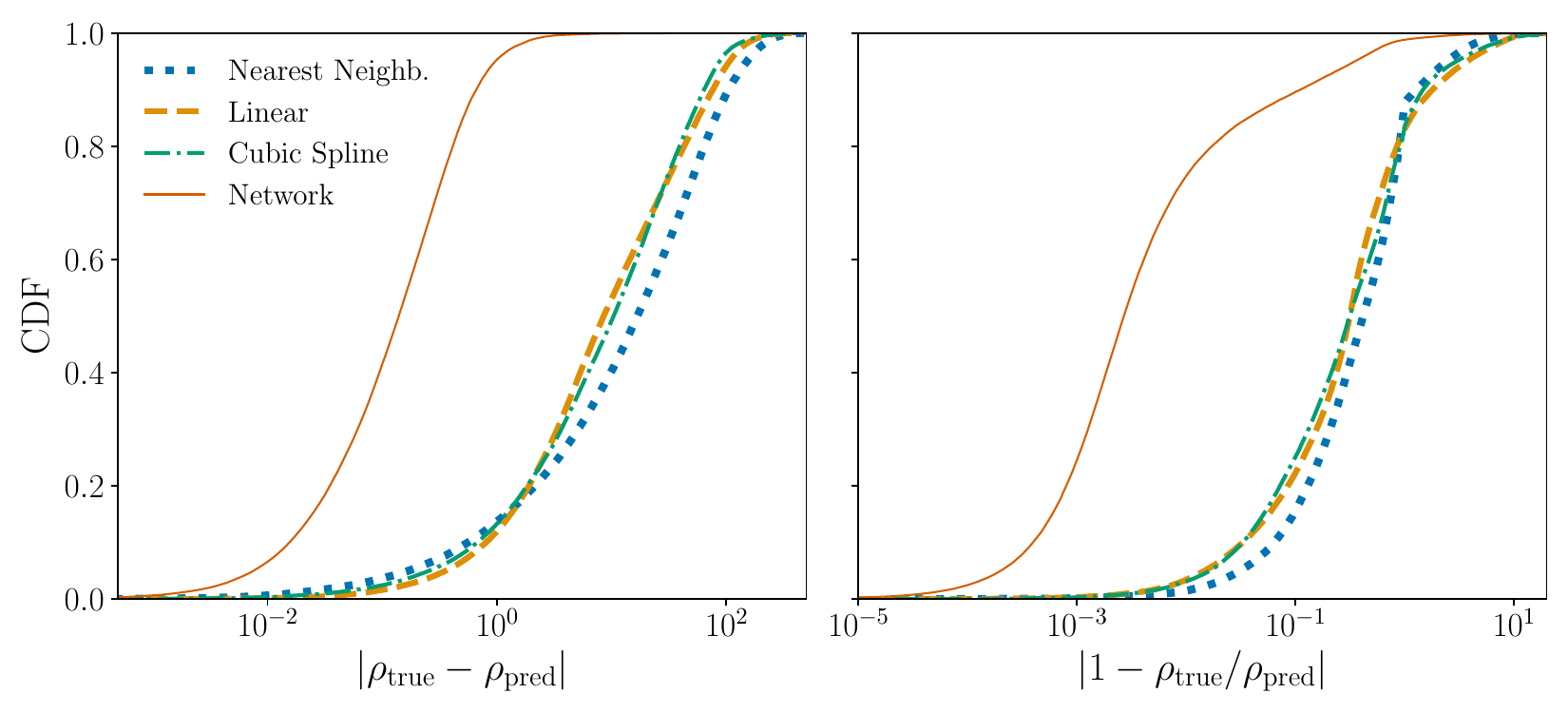}
    \caption{Cumulative distribution functions for the (\emph{left}) absolute and (\emph{right}) \changed{fractional} differences between the SNR predictions of (linear, nearest neighbour, cubic spline, and neural network) interpolation methods and the true values. 
    The former three grid-based interpolation methods perform poorly, with at least $47\%$ of SNRs inaccurate by at least $10$, and at least $75\%$ predicted to less than $10\%$ accuracy. 
    Conversely, the latter method precisely estimates SNRs: $95.3\%$ of SNRs are determined to within $1$, and $89.6\%$ within $10\%$ of the true values.}
    \label{fig:interps}
\end{figure*}

Neural networks are highly flexible mathematical tools that are capable of learning complex relationships in high-dimensional spaces \citep{Goodfellow2016}.
For our purposes, we need a neural network that takes a vector as an input (the EMRI parameters) and produces a scalar output (the SNR estimate). 
We opt for the multilayer perceptron (MLP) algorithm \citep{Hastie01} as it fits this specification.
MLPs are fast and capable of high accuracy, satisfying our requirements well.
The design and training of this MLP are discussed in Appendix~\ref{sec:nn_design}.
The trained network achieves two orders of magnitude of improvement in accuracy compared to other interpolation approaches, as shown in Figure \ref{fig:interps}; the majority of the test data are predicted to percent-level accuracy.
This network is capable of producing $10^5$ SNR estimates in $< 0.1~\mathrm{s}$, which is six orders of magnitude faster than calculating the SNR directly.

As the MLP was trained with an L1 loss function, which minimises the absolute difference between the prediction and truth \citep{Goodfellow2016}, it does not perform as well \changed{in terms of fractional error} for $\rho_\mathrm{opt} \ll 1$.
This manifests as a larger upper tail in the relative CDF.
However, this does not translate to a reduction in performance, as these signals are too weak to be detectable across the majority of the luminosity distance distribution.
If adequate performance across all SNR scales is required, this may be achieved with the appropriate choice of loss function\changed{, for instance by} training on the log of the SNR.
By choosing \changed{not to train on log SNRs}, we prioritise the regions of parameter space corresponding to larger SNRs in the data set (at the fiducial luminosity distance of $1~\mathrm{Gpc}$).
As these SNRs will be pushed towards the detection threshold at larger distances, and the majority of our luminosity distance distribution is above $1~\mathrm{Gpc}$, estimating these larger SNRs well has the greatest impact on accurate detectability estimates.

\subsection{Interpolating over the selection function}
\label{sec:MLP2}

Using our trained interpolator network to produce SNR estimates, we are now able to approximate $\alpha(\bm{\lambda})$ via Eq.~\eqref{eq:alpha_eq2} with the interpolated SNR as an input. 
However, two issues still remain that must be addressed before this may be applied in practice.
First, this setup still requires that the parameter set $\{\Vec{\theta}\}$ is drawn from  $p_\mathrm{pop}(\Vec{\theta}|\Vec{\lambda})$ for each hyperlikelihood call; this is a slow operation, even for the relatively simple population models in Table~\ref{tab:populations}. 
It is also inefficient, because $\alpha(\bm{\lambda})$ will typically not vary significantly across the high-probability region of the posterior, where the majority of samples are drawn.
Second, the stochastic nature of the Monte Carlo selection function estimates itself presents challenges in sampling: the hyperlikelihood surface becomes noisy, which can be problematic for the reliable convergence of sampling algorithms.
These issues prohibit the use of our stochastic selection function estimates in hyperposterior sampling and must be solved.

To address these problems and further accelerate our bias-corrected likelihood, we extend the idea of interpolating over high-dimensional spaces with MLPs further. A second MLP trained prior to sampling can be used to interpolate directly over $\alpha(\bm{\lambda})$.
The architecture and training settings for this MLP are discussed in Appendix~\ref{sec:nn_design}.
Using this second MLP step greatly reduces the time per likelihood call, achieving a further order of magnitude of speedup with respect to stochastic estimation of the selection function. 
For our chosen population model, the resulting computational cost of the numerator and denominator in Eq.~\eqref{eq:hyperposterior} become roughly equivalent.
Further speedup is achieved with vectorised evaluation of the hyperlikelihood, as the selection function MLP is capable of handling many sets of hyperparameters at no additional cost provided that sufficient GPU memory is available. 
This vectorisation would not be practical if one were to estimate the selection function values with the stochastic approach.
With the main limitations of our method addressed, we are now able to produce selection function estimates usable in sampling at low computational cost.

\section{Unbiased population inference results}
\label{sec:sampling}

With our strategy for selection bias correction formulated, we are now properly equipped to tackle an EMRI population inference problem.
We begin by simulating a catalogue of EMRI observations.
To estimate how well \textit{LISA} will resolve the shape of the EMRI population in a realistic scenario, we choose hyperparameter values supported by recent black hole population studies:
\begin{itemize}
    \item 
    While constraints have been placed on the slope of the MBH mass function by current observations \citep{Shankar2013}, the mass function for MBHs hosting EMRIs is subject to additional selection effects that are poorly understood at present \citep{Babak2017}.
    Recent work has estimated that the spectral index of this power law after the inclusion of selection effects is $\lambda_M \approx -1.43$ in the mass range $[M_\mathrm{min},M_\mathrm{max}] = [10^5,10^7]~M_\odot$ \citep{Babak2017}.
    We assume that the slope of the CO mass function is equal to the median value observed in stellar-mass BBH mergers of $\lambda_\mu \approx -3.50 $ in the mass range $[3,90]~M_\odot$ \citep{o3pop}, assuming that the progenitors of these mergers are representative of the universal stellar-mass BH population.
    The selection effects that translate this into the EMRI CO mass function are poorly understood and would require a dedicated set of $N$-body simulations of stellar cusps to properly quantify \citep{Broggi2022, Babak2017}, so we do not consider them in this study.
    Despite this caveat, the slope observed via BBH mergers is the strongest constraint placed on the mass function for black holes in this mass range available and is a reasonable starting point for estimating \textit{LISA}'s ability to resolve the CO mass function with EMRI observations.
    \item
    The MBH spin magnitude distribution is also poorly constrained by observational data. 
    Current measurements are limited to MBHs in active galactic nuclei \citep{Daly2011}, which may not be representative of the full MBH spin magnitude population as different formation channels will yield different MBH spin magnitude distributions \citep{Amaro-Seoane2022}.
    Self-consistent simulations of MBH growth with cosmic evolution predict that most MBHs have spins greater than $0.9$ in the MBH mass range quoted above, with a fairly narrow spread below $10^7~M_\odot$ \citep{Sesana2014}.
    We include this characteristic shape of a narrow spin distribution above $a=0.9$ in our population by choosing $[\mu_a, \sigma_a]=[0.93, 0.06]$.
    \item
    The time that each EMRI plunges with respect to the start of the observational data $t_\mathrm{plunge}$ is randomly distributed in the range $[0,10]~\mathrm{yr}$.
    In line with the planned \textit{LISA} mission duration, we assume a 4-year observational window \citep{Amaro-Seoane2017}: some EMRIs will not plunge until after the end of our observational data, but may still be detectable if they are bright enough). 
    We assume that EMRIs occur at a rate of $\mathcal{R}=700~\mathrm{yr}^{-1}$, which is conservative (considering our redshift cut-off of $z=6$) when compared with EMRI rate estimates from astrophysical modelling \citep{Babak2017, Broggi2022, Vazquez-Aceves2021}.
\end{itemize}
After discarding the signals too faint to be detected, we obtain a catalogue of $116$ EMRIs.

To demonstrate the selection biases present, we perform two sampling runs: one in which selection biases are corrected for with our interpolation scheme, and another in which selection effects are not accounted for, i.e., replacing $\alpha(\Vec{\lambda})$ with $1$ in Eq.~\eqref{eq:hp_likelihood} and Eq.~\eqref{eq:rate_likelihood}.
We sample the hyperposterior Eq.~\eqref{eq:hyperposterior} with the \texttt{nessai} nested sampler \citep{Williams2021, nessai}, using default settings.
The convergence of all sampling runs with these settings was verified by examining the results of internal consistency checks built into \texttt{nessai}.
The hyperposteriors obtained from these sampling runs for a subset of hyperparameters are shown in Figure~\ref{fig:joint_corner_subset}. 
The full hyperposterior is shown in Appendix~\ref{sec:fullCorner}, which demonstrates minor discrepancies between the two hyperposteriors for the other hyperparameters, with marginal posteriors that are too narrow (over-constrained) but otherwise fairly consistent with the set hyperparameter values.

The bias that results from ignoring selection effects on the observations is visible here as a discrepancy between the credible interval contours of the two hyperposteriors at the $99\%$ level.
For $\lambda_M$, the uncorrected posterior is inconsistent with the true value at the $99\%$ credible level; the marginal posteriors for $\lambda_\mu$ that include or exclude selection bias correction disagree to an even greater extent.
There is also a clear difference between the marginal posteriors on $\mathcal{R}$: this is symptomatic of the presence of selection effects, as it indicates that predicting the overall event rate solely from the size of the detection catalogue will result in a significant underestimate of the actual event rate.

After accounting for selection effects, our results serve to probe how well \textit{LISA} can resolve the form of this EMRI population.
We estimate (quoting the median and the $90\%$ credible interval) that $\lambda_M = -1.39 ^{+0.12}_{-0.12}$ and $\lambda_\mu = -3.58 ^{+0.16}_{-0.17}$, corresponding to precisions of $8.8\%$ and $4.6\%$ respectively.
The MBH spin distribution is well recovered, with $\mu_a = 0.924^{+0.008}_{-0.007}$ and $\sigma_a = 0.054^{+0.006}_{-0.005}$; these hyperparameters are recovered to within $0.87\%$ and $10\%$ respectively.
The EMRI rate is estimated with $12\%$ precision to be $\mathcal{R} = 678^{+88}_{-75}~\mathrm{yr}^{-1}$.
The precision achievable by \textit{LISA} will roughly scale with the square root of the true event rate, which we have assumed a conservative value for in this study: for the most optimistic scenarios, this could improve by as much as an order of magnitude \citep{Babak2017}. 

\changed{The number of detected events depends on the underlying population. 
For example, if the number of events is skewed to high redshift with respect to our assumed distribution of uniform in comoving volume and comoving time, then the number of detected events will decrease accordingly.
Our choice of redshift distribution is equivalent to assuming that the probability of an EMRI occurring for a given MBH is constant across cosmic time.
In reality, we expect that the physics of EMRI formation, such as cusp erosion \citep{Babak2017}, will lead to deviations away from this. 
Similarly, we expect that the distributions of MBH and CO masses will differ in reality from our assumptions. 
Hence, the results presented here should only be considered illustrative. 
A comprehensive study of how population inference results vary with the underlying population (which would require a computationally efficient method, such as ours) is necessary to fully map out how well \textit{LISA} could measure the EMRI source population.}

\begin{table}
	\centering
	\caption{Hyperprior distributions chosen for all sampling runs. As the range of plunge times is reduced by a factor of 5 for our probability--probability plot analysis (Section~\ref{sec:pp}), our prior bounds on the EMRI rate are adjusted accordingly: this is indicated by $(^*)$.}
	\label{tab:priors}
	\begin{tabular}{ccc} 
		\hline
		Parameter & Distribution & Limits\\
		\hline
		$\lambda_M$ & Uniform & $[-4, -1]$\\
		$M_\mathrm{min}$ & Uniform & $[5, 50]\times 10^4~M_\odot$\\
		$M_\mathrm{max}$ & Uniform & $[5, 50]\times 10^6~M_\odot$\\
		$\lambda_\mu$ & Uniform & $[-4, 1]$\\
		$\mu_\mathrm{min}$ & Uniform & $[1, 5]~M_\odot$\\
		$\mu_\mathrm{max}$ & Uniform & $[80, 100]~M_\odot$\\
		$\mu_a$ & Uniform & $[0.05, 0.95]$\\
		$\sigma_a$ & Uniform & $[10^{-3}, 2]$\\
		$\mathcal{R}$ & Log-uniform & $[350, 1050]\  ([75,150]^*$)$~\mathrm{yr}^{-1}$ \\
		\hline
	\end{tabular}
\end{table}

\begin{figure}
	\includegraphics[width=\columnwidth]{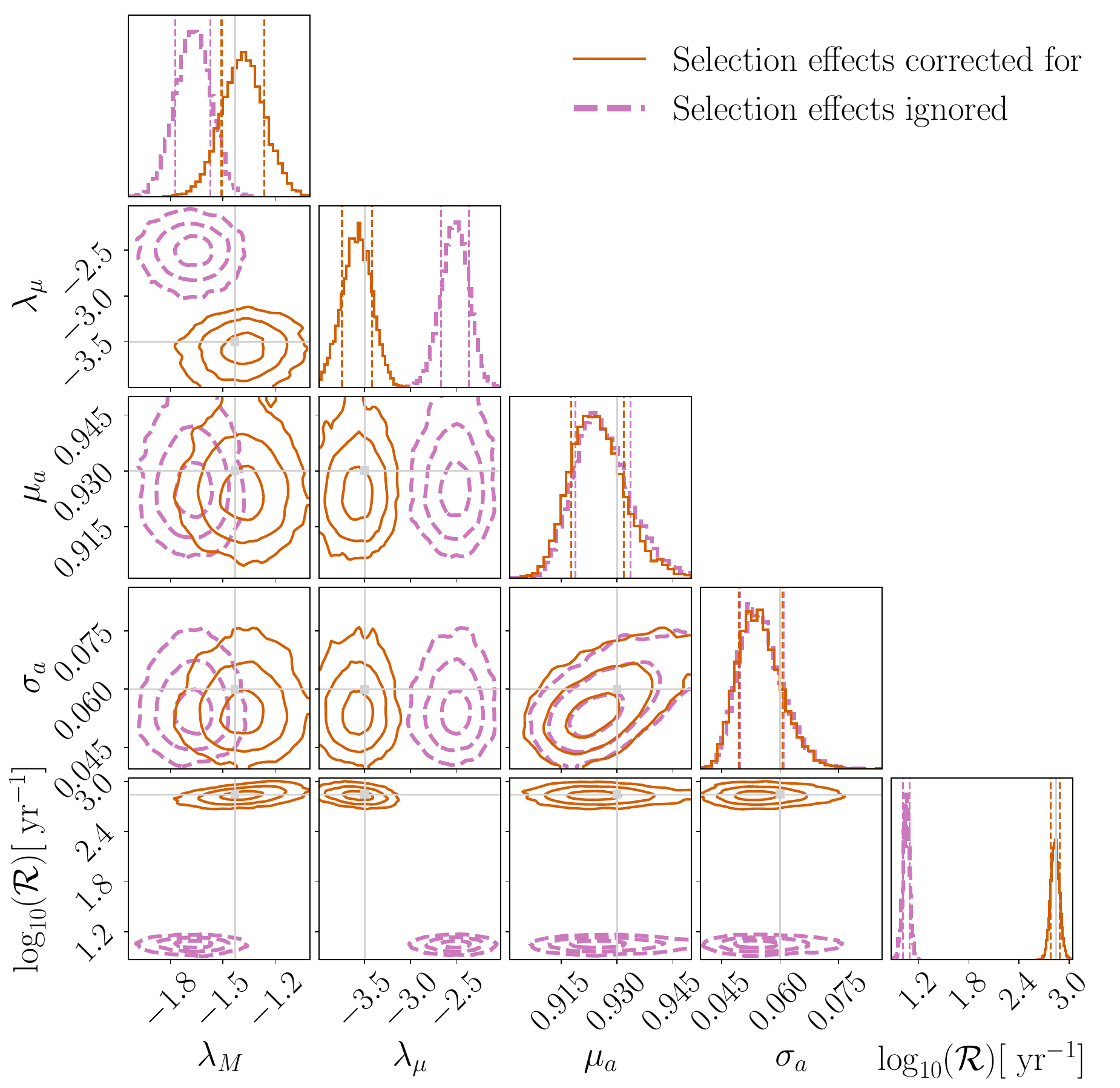}
    \caption{Recovered hyperposteriors for our example scenario with selection effects corrected for (solid) or ignored (dashed). 
    The true values of each hyperparameter are indicated by the cross-hairs. 
    Significant bias is present in the recovery of $\lambda_\mu$, with minor bias also visible in the recovery of $\lambda_M$. 
    The rate $\mathcal{R}$ is accurately recovered with the inclusion of selection effect correction.}
    \label{fig:joint_corner_subset}
\end{figure}

\section{Verifying the accuracy of results}
\label{sec:pp}

In the previous section, we demonstrated the capability of our approach for a single example. 
However, this is not sufficient to ensure that the selection function estimates output by our MLP are sufficiently accurate and unbiased that population inference will return hyperposteriors that are consistent with the truth across the hyperparameter space. 

To assess whether this is the case, we opt for the probability--probability (P--P) plot test \citep{Cook2006}.
First, we draw $N$ sets of hyperparameters from the hyperpriors described in Table~\ref{tab:priors}, and generate the corresponding population catalogues.
We then perform hyperposterior sampling runs to produce estimates of the hyperparameters in each case, and determine the confidence interval $q_\lambda$ of the true hyperparameters with respect to the posterior obtained. 
Last, we plot the CDF of $q_\lambda$. 
When the trial sets of hyperparameters are drawn from the hyperprior, we expect that the true value of a hyperparameter will fall within the $x\%$ credible interval in $x\%$ of realizations (i.e., a plot of $q_\lambda$ against its CDF will be diagonal) if our hyperposteriors are consistent with the true values in all cases. 
We test the accuracy of our inference framework by comparing the calculated CDF with the expected diagonal trend.

\begin{figure}
	\includegraphics[width=0.94\columnwidth]{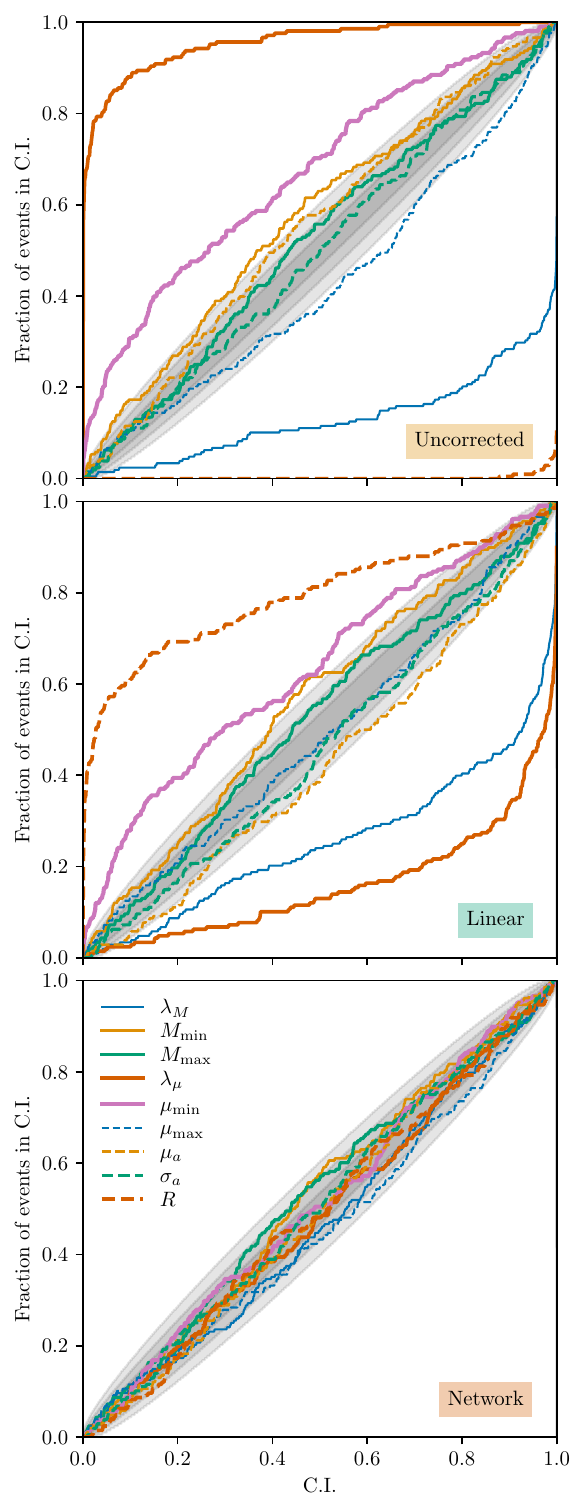}
    \caption{Probability--probability plots showing the fraction of hyperparameters within a given credible interval (CI) for $N=208$ hyperprior draws. 
    Leaving selection biases uncorrected, setting $\alpha(\Vec{\lambda}) = 1$, demonstrates the presence of significant selection biases in the population inference (\emph{top}). 
    These biases are successfully rectified with our approach (\emph{bottom}), whereas standard interpolation techniques fail to produce unbiased results (\emph{middle}).
    The expected $68\%$, $90\%$ and $99\%$ confidence intervals are shown in grey.}
    \label{fig:pp}
\end{figure}

Some variation of each CDF from the diagonal due to small-number statistics is expected. 
For the P--P plot to be meaningful $N$ needs to be large, so we modify our population to reduce the computational cost of waveform generation by reducing the length of the observational window from $4~\mathrm{yr}$ to $2~\mathrm{yr}$ and the range of EMRI plunge times from $[0, 10]~\mathrm{yr}$ to $[0,2]~\mathrm{yr}$. 
To further reduce the cost of generating each population, we lower the overall event rate by limiting our population to a maximum redshift of $z=1$.
Adjusting the event rate to account for these reductions in both duration and sensitive volume, the number of expected EMRIs for each population decreases by a factor of $35$. 
The simulation and analysis configuration remains otherwise unchanged from the analysis described in Section~\ref{sec:sampling}.
We perform three analyses to compare their results: first, we exclude selection effect correction; second, we include selection effect correction by means of a linear interpolation scheme, and last, we include selection effect correction with our neural network interpolation scheme.
The P--P plots obtained from these analyses are shown in Figure~\ref{fig:pp}, broken down by hyperparameter and compared to the expected $68\%, 90\%$ and $99\%$ deviations for the $N=208$ sets of drawn hyperparameters \citep{Ibe2013}.

We first examine the P--P plot for the hyperposteriors obtained when selection effects are excluded, setting $\alpha(\Vec{\lambda}) = 1$, as shown in the top panel of Figure~\ref{fig:pp}. 
The resulting posteriors exhibit strong biases with respect to the true values, and the P--P test fails; the majority of the hyperparameter CDFs deviate beyond the $99\%$ confidence interval.

The P--P plot is a useful diagnostic for how the posterior deviates from what is expected on a dimension-by-dimension basis.
We can identify that $\lambda_\mu, m_\mathrm{min}, \mu_\mathrm{min}$ and $\mu_a$ are typically overestimated when selection effects are ignored, and conversely that $\mu_\mathrm{max}$ and $\lambda_M$ are underestimated. 
This can be understood by considering how the SNR of an EMRI waveform is correlated with the parameters associated with these hyperparameters. 
The SNR is positively correlated with $\mu$, so the power-law slope is flattened off by the suppression of low CO mass events; similarly, SNR is higher for larger MBH spins, so the mean of the truncated Gaussian shifts to the right due to the suppression of the lower spin events. 
For $\lambda_M$, the opposite is true: the number of detectable higher MBH mass EMRIs (which are at lower frequencies) is suppressed due to the shape of the \textit{LISA} sensitivity curve and so the observed power-law steepens. 
The parameters that pass the P--P test, $M_\mathrm{max}$ and $\sigma_a$ are not strongly influenced by selection bias. 
The mean of the MBH spin truncated Gaussian may shift, but the change in the width of the distribution will be proportionally smaller, and is therefore less sensitive to this selection effect.
Likewise, as the high MBH mass EMRIs are typically unobserved and occupy a small fraction of the overall EMRI population, small changes to the maximum MBH mass do not strongly affect the detectable fraction of the population.
This is not the case for the CO mass distribution: the high mass events in the upper end of the power law are also the brightest events in the population, so adjusting the upper limit of the power law leads to larger changes in the fraction of events expected to be detectable.
As the observed deviations from consistent hyperposteriors align with our expectations, we are confident they are the result of selection effects.

To demonstrate the impact of the low interpolation accuracy seen in Figure~\ref{fig:interps} on the resulting selection function estimates, we repeat this analysis with an MLP selection function trained on SNR estimates produced by the linear interpolator described in Section~\ref{sec:snr_interp}. 
The resulting P--P plot, shown in the middle panel of Figure~\ref{fig:pp}, demonstrates that although modest correction is achieved in this case, it is still far too biased and inaccurate to result in consistent hyperposteriors. 
In the case of $\lambda_\mu$, this even results in an over-correction of the selection bias when compared to the uncorrected case.
The danger of over-correction implies that it is not sufficient to include a selection function term in the population likelihood: the selection function must also be accurately calculated to obtain good results.

Finally, we present the P--P plot obtained for this analysis with our MLP selection function estimator in the bottom panel of Figure~\ref{fig:pp}.
In stark contrast to the previous two plots, the hyperparameter CDFs are fully consistent with the expected confidence intervals.
This conclusion is supported by the results of Kolmogorov--Smirnov \citep{Dodge2008} tests for each hyperparameter.
Combining the p-values from each test with Fisher's method \citep{Mosteller1948} returned a combined p-value of $0.3$.
This \changed{indicates} that all hyperparameters (including the EMRI rate $\mathcal{R}$) are consistently recovered: the MLP is capable of producing selection function estimates that are sufficiently accurate for consistent posteriors to be obtained.
This result verifies the application of our method in the treatment of selection biases in population inference.

\section{Conclusions}
 
Population inference with EMRIs has the potential to probe the evolution of both MBHs and their galactic neighbourhoods to unprecedented precision.
However, the computational cost of components of this analysis is prohibitively high.
Estimating selection biases in EMRI populations is computationally expensive due to a combination of the need for costly waveform models and the resources required to perform SNR calculations for long-duration data. 
As the SNR calculation is the computational bottleneck, we substitute it for an interpolation over pre-computed SNRs.
We find that commonly employed interpolation schemes are not sufficiently fast or accurate for this problem, so we instead use machine-learning techniques.
Using a neural network trained on a data set of SNRs distributed uniformly in the EMRI parameter space, we achieve a speedup of six orders of magnitude over direct SNR evaluation.
We then replace the stochastic estimation of the selection function obtained via Monte Carlo integration with a second neural network that is trained on these stochastic estimates distributed uniformly in the hyperparameter space.
This further improves the speed of hyperlikelihood evaluation by an order of magnitude, and enables vectorised estimation of the selection function for further reductions in computational cost.
To verify the robustness of our approach against systematic biases, we globally evaluate hyperposterior consistency by simulating $208$ EMRI populations and checking the results of selection bias-corrected hyperparameter estimation with a P--P test.
This test confirmed that (i) the presence of selection effects significantly biased inferences that did not correct for them appropriately, and (ii) our approach successfully corrected for selection effects to produce unbiased results.

We apply our method to the inference of an astrophysically-motivated EMRI population \changed{(assuming sources distributed uniformly in comoving volume and time)} to study \textit{LISA}'s ability to probe the structure of such populations.
We estimate that $\lambda_M = -1.39 ^{+0.12}_{-0.12}$ (a precision of $8.8\%$) and that $\lambda_\mu = -3.58 ^{+0.16}_{-0.17}$ (a precision of $4.6\%$).
For the MBH spin magnitude distribution, we find that $\mu_a = 0.924^{+0.008}_{-0.007}$ and that $\sigma_a = 0.054^{+0.006}_{-0.005}$, resolving the width of the MBH spin magnitude distribution to within $10\%$.
The event rate is estimated to be $\mathcal{R} = 678^{+88}_{-75}~\mathrm{yr}^{-1}$ (a precision of $12.0\%$).

The capability of our approach for treating selection effects in the case of a simple population model, which excludes substructure or correlations due to astrophysical effects, paves the way for future work to investigate more complex EMRI population models.
The EMRI population we expect to be present in reality is multi-faceted \citep{Babak2017}, and by introducing these features systematically to the population inference problem we can begin to characterise their measurability with space-based detectors such as \textit{LISA}. 
As population inference is intimately tied to other hierarchical inference problems, including cosmological inference \citep{MacLeod2008, Laghi2021} and tests of general relativity \citep{Chua2018}, proper treatment of selection effects for EMRIs has direct implications for these analyses as well.
Ultimately, a joint hierarchical inference over this problem space may be required, of which our approach can be an integral part.

As our method is not predicated on a particular population model, it is applicable to a wide variety of population inference problems.
Similarly, as any waveform model may be used, this approach is capable of accommodating future changes to EMRI waveform models with little tuning required. 
While our approach specifically targets the EMRI population inference problem, it may be generalised to any problem with an SNR-threshold selection bias due to this model-agnostic nature. 
The reduction in computational cost achieved by employing our method will be most pronounced in cases where the SNR function is expensive to compute and of high dimensionality, but the ability to form a vectorised SNR approximant will still offer a notable speedup for waveform models that are not easily parallelisable or vectorisable (e.g., due to memory constraints).

\changed{Our code package \texttt{poplar}, containing the tools used in this paper, has been released as an open-source package \citep{poplar} for use in future population studies.}

\section*{Acknowledgements}
We thank Michael Katz and Alvin Chua for useful discussions regarding EMRI waveforms and detection, \changed{and Alvin Chua for additional careful review of the manuscript}.
CEAC-B thanks Michael Williams for useful discussions on parameter estimation, sampling and good coding practices.
This work was performed using \texttt{numpy} \citep{numpy}, \texttt{cupy} \citep{cupy_learningsys2017} and \texttt{PyTorch} \citep{PyTorch}.
Figures were produced using \texttt{matplotlib} \citep{matplotlib}, \texttt{seaborn} \citep{seaborn} and \texttt{corner} \citep{corner}.
This work makes use of the Black Hole Perturbation Toolkit.\footnote{\href{http://bhptoolkit.org/}{bhptoolkit.org/}}
CEAC-B is supported by STFC studentship 2446638. 
This work was supported in part by STFC grant ST/V005634/1.

\section*{Data Availability}
This paper is accompanied by a tagged data release \citep{Chapman-Bird2022}, consisting of: (i) the data and scripts required to reproduce the figures presented; (ii) the trained neural networks used to produce the results presented, along with the code required to use them; (iii) scripts for generating the data used to train the neural networks.

\bibliographystyle{mnras}
\bibliography{example}

\appendix

\section{EMRI parameter estimation and catalogue generation}
\label{sec:IndivEMRIs}
The generation of a catalogue of EMRI observations is multi-faceted.
We first describe the generation of EMRI waveforms, including our treatment of initial conditions and our choices regarding waveform model and detector response, in Section \ref{sec:EMRIwaveforms}.
In Section \ref{sec:EMRIpe_appendix}, we outline our procedure for obtaining posterior samples for EMRIs that pass the detection threshold, along with consistency checks we perform to ensure the approximations made in this process are justifiable. 
These two steps are performed for each set of EMRI parameters drawn from the population to construct a catalogue of EMRI detections and their corresponding posterior samples.

\subsection{Parametric conventions and waveform generation}
\label{sec:EMRIwaveforms}

The EMRI parameter space is complicated, consisting of $18$ parameters:
\begin{itemize}
    \item 
    The intrinsic parameters of the EMRI describe the properties of the two objects and their initial orbital configuration.
    The primary MBH is described by its mass $M$ and spin vector $\Vec{a}$, and the secondary CO similarly by mass $\mu$ and spin vector $\Vec{a}_\mathrm{CO}$. 
    As $\Vec{a}_\mathrm{CO}$ is not predicted to have a \changed{significant} effect on EMRI \changed{detectability} \citep{Huerta2011}, it is not currently included in state-of-the-art waveform models, including the model used in this study \citep{Katz2021}.
    The inspiral orbit is described by the initial eccentricity $e_0$, semi-latus rectum $p_0$ and orbital inclination $\iota_0$, along with three orbital phases $\Phi_r,\Phi_\theta$ and $\Phi_\phi$ \citep{Fujita2020}. 
    
    \item 
    The extrinsic parameters describe the orientation of the system and its location with respect to the detector.
    The position of the system is described by the luminosity distance vector $\Vec{d}_\mathrm{L}$, the magnitude of which may be described by a redshift $z$ via the relation
    \begin{equation}
        d_\mathrm{L} = (1+z)\frac{c}{H_0}\int_0^z \frac{\mathrm{d}z'}{E(z')},
    \end{equation}
    where $H_0$ is the Hubble constant \citep{Hogg1999}.
    We separate out vectors into their magnitudes and angular components, such that $\Vec{d}_\mathrm{L} = \{d_\mathrm{L},\theta_\mathrm{S},\phi_\mathrm{S}\}$. 
    Similarly, we decompose the primary spin magnitude vector $\Vec{a} = \{a,\theta_\mathrm{K},\phi_\mathrm{K}\}$.
    The sets of angles describe the orientation of the MBH spin-angular momentum vector and the sky position vector respectively, with $\theta$ and $\phi$ referring to polar and azimuthal angles respectively.
    \item
    The plunge time of the system with respect to the start of the observation is described by the parameter $t_\mathrm{plunge}$.
    However, the concept of initial conditions on a population level is not well-defined unless a common reference point in the waveform is set.
    Therefore, we adjust $p_0$ such that the EMRI waveform will plunge after $10~\mathrm{yr}$ \citep{Stein2020}. 
    For our fiducial example outlined in Section~\ref{sec:sampling}, we assume a \textit{LISA} observing window of $4~\mathrm{yr}$ in line with the current mission proposal \citep{Amaro-Seoane2017}. 
    Our choice to allow for EMRI plunges to occur up to $10~\mathrm{yr}$ after the beginning of \textit{LISA} observation is made to accommodate the presence of EMRIs in the data that plunge after the end of the observation window. 
    Neglecting these events as being undetectable is not typically a reasonable approximation. 
    In many cases, $\rho_\mathrm{opt} > \rho_\mathrm{t}$ even for $t_\mathrm{plunge} = 10~\mathrm{yr}$, so some detectable events in the data are ignored despite this extension of EMRI plunge times to the post-window regime. 
    By excluding these events that plunge after $10~\mathrm{yr}$, the results of our population inference will be conservative.
    We include this effect to demonstrate that our approach is capable of accommodating a post-window cut on $t_\mathrm{plunge}$, but acknowledge that the tuning of such a cut-off point with respect to the detectability of the excluded signals is an issue that warrants further investigation in future work.
\end{itemize}

Waveform generation also includes some additional considerations to transform from the source frame to the detector frame.
Prior to waveform generation, we convert source-frame masses to detector-frame masses with the mapping $M_\mathrm{det} = (1+z)M$ \citep{Krolak1987}.
For convenience, we do not include a detector response in our EMRI waveform modelling and instead work in terms of the waveform strain (as opposed to the TDI combinations that the \textit{LISA} detector outputs will be used to construct \citep{Tinto2021}). 
This choice leads to a degeneracy between $\phi_\mathrm{K}$ and $\phi_\mathrm{S}$, which we navigate by defining a new parameter $\Delta\phi=\phi_\mathrm{S} - \phi_\mathrm{K}$. 
Our approach can incorporate a chosen \textit{LISA} response by reverting back to the separate angles and including the response function in waveform generation.
As the addition of a response function does not significantly alter EMRI SNRs, we do not expect its exclusion to affect the validity of our approach.

For our waveform model, we choose the fifth-order post-Newtonian Augmented Analytic Kludge recently implemented in the FEW package \citep{Katz2021, Chua2017}. 
The validity of our population inference framework should not depend strongly on this choice, as the EMRI SNR function should remain well-behaved and smooth for any reasonable choice of waveform model, although the specific numerical results may vary for different waveforms. 

To accommodate our choice of initial conditions, we generate $10~\mathrm{yr}$ EMRI waveforms in the time domain with a sampling rate of $0.1~\mathrm{Hz}$, and crop them according to their (randomly sampled) $t_\mathrm{plunge}$ values.
We calculate the waveform's $\rho_\mathrm{opt}$ value via Eq.~\eqref{eq:opt_snr} and produce a noise-realized SNR estimate $\rho_\mathrm{n}$ by drawing a sample from Eq.~\eqref{eq:simp_chi2}. 
Detection is evaluated via Eq.~\eqref{eq:alpha_eq1}.

\subsection{Parameter estimation}
\label{sec:EMRIpe_appendix}
For waveforms that pass the detection threshold, we proceed to draw samples from the posterior distribution on $\Vec{\theta}$. 
Operating under the LSA, we determine the FIM $\Gamma$ from numerical waveform derivatives computed using the five-point stencil method \citep{Sauer12}. 
Appropriate step sizes that produce accurate (and stable) numerical derivatives were determined empirically by computing $\langle \partial_\ell h | \partial_\ell h \rangle$ (the FIM diagonal terms) on a grid and identifying regions of step-size space for which this converged.
This stability was then verified across the EMRI parameter space.
With $\Gamma$ computed, we then invert it to obtain the covariance matrix of the LSA likelihood. 
FIMs for EMRI waveforms typically have large condition numbers, which can cause issues when performing matrix inversion; we mitigate this by employing singular value decomposition (SVD) to compute the pseudoinverse of $\Gamma$ \citep{Ben-Israel2003}. 
The numerical stability of this inversion can also be problematic, even for double precision; to alleviate this, we perform the SVD with $500$-point decimal precision using the \texttt{mpmath} package \citep{mpmath}.
We also perform some additional verification of the validity of the LSA likelihood as a substitute for the full likelihood as recommended in \citet{Vallisneri2008}. 
This check is performed by examining the ratio between the LSA and full likelihood, which we denote $r$. 
By sampling the $1$ standard deviation isoprobability contour in the LSA likelihood and computing $|\log_{10} r|$ for each of these samples, we can examine whether the high-probability regions of the LSA and the full likelihood are consistent.
Following \citet{Vallisneri2008}, we accept the LSA likelihood as a suitable approximation if and only if $90\%$ of the samples satisfy $|\log_{10} r| < 0.1$. 
We found that for waveforms with $\rho_\mathrm{opt} > 10$ that this condition was satisfied in all cases, which is well below our chosen SNR threshold of $20$.

Our goal is to obtain posterior samples via the LSA that are representative of those that would be obtained from sampling the true posterior.
In the absence of noise, the likelihood will peak on the true values provided there are no degeneracies in the parameter space.
For EMRIs, this is not generally true, but the degeneracies are non-local and therefore do not affect the morphology of the posterior near the true values \citep{Chua2022}.
However, when noise is included, the likelihood shifts in a random direction in parameter space such that the true values are no longer at the maximum likelihood point.
To simulate this measurement effect, we draw one sample from the posterior (which, assuming broad uniform priors, is equivalent to the event likelihood) and recompute the FIM at this new point \citep{Simon2017}. 
Finally, with our second FIM computed, we invert it and draw $S=10^4$ samples from the corresponding multivariate normal distribution. 
The posterior samples in $d_\mathrm{L}$ are converted to samples in $z$ assuming a standard cosmology, and detector-frame masses then converted back to the source frame by dividing through by $(1+z)$ on a per-sample basis.
This modified set of posterior samples is the final product of the individual EMRI event simulation.

\section{Neural network design and training}
\label{sec:nn_design}

MLPs are tuneable mappings between input and output vectors (of specified lengths) that consist of a layered structure of matrix multiplications which are passed through non-linear functions.
The non-linearity between each layer, combined with a large number of tuneable parameters in each matrix multiplication, enables the resulting neural network to mimic complex mappings between high-dimensional spaces \citep{Goodfellow2016}.
This tuning is performed in a process known as training, in which the performance of the neural network is maximised with respect to a pre-computed data set.

The number of required neurons and layers (which describe the dimensions of each matrix multiplication) in an MLP depends on the complexity of the function to be interpolated and the number of interpolation dimensions.
Due to the stochastic nature of training neural networks, tuning of the learning rate, batch size and number of iterations employed during training is required to maximise performance.
For complicated problems, optimization techniques may be employed to explore the space of network settings and identify a sensible configuration \citep{Feurer2019}.
In our case, the problem is sufficiently low-dimensional that we were able to obtain effective MLPs through the manual tuning of network settings.
Network complexity was gradually increased through the addition of neurons or layers until overfitting \citep{Goodfellow2016} was observed. 
This is characterised by the performance of the network on testing data degrading despite continued improvement in performance on training data.
At this point, training settings were adjusted to minimise this overfitting.
The choice to rescale training data to that of a unit normal distribution, and to use the Adam optimisation algorithm \citep{Kingma2014}, was made following current best practices \citep{Goodfellow2016}: use of other optimisation or rescaling functions was not found to significantly affect network performance.

The resulting network settings chosen for the SNR and selection function MLPs, summarised in Table~\ref{tab:mlp_settings}, are almost identical.
Two extra hidden layers are added for the SNR MLP, which is to be expected given the higher dimensionality of the EMRI parameter space in comparison to the hyperparameter space.

\begin{table}
    \centering
    \caption{Architecture and training settings chosen for our two multilayer perceptron (MLP) neural networks introduced in Section~\ref{sec:snr_interp}.}
    \label{tab:mlp_settings}
    \begin{tabular}{ccc}
    \hline
    Setting & SNR MLP & $\alpha(\lambda)$ MLP\\
    \hline\hline
    Number of (hidden) layers & $10$ & $8$\\
    Neurons per layer & $128$ & $128$\\
    Activation function & SiLU & SiLU\\
    Rescaling & Unit normal & Unit normal\\
    \hline
    Optimiser & Adam & Adam\\
    Learning rate & $5 \times 10^{-4}$ & $5 \times 10^{-4}$\\
    Batch size & $10^4$ & $10^5$\\
    Max epochs & $10^5$ & $10^3$\\
    Loss function & L1 & L1 \\
    \hline
    \end{tabular}
\end{table}

\section{Full hyperposterior obtained from 4-year scenario}
\label{sec:fullCorner}

\begin{figure*}
	\includegraphics[width=0.8\paperwidth]{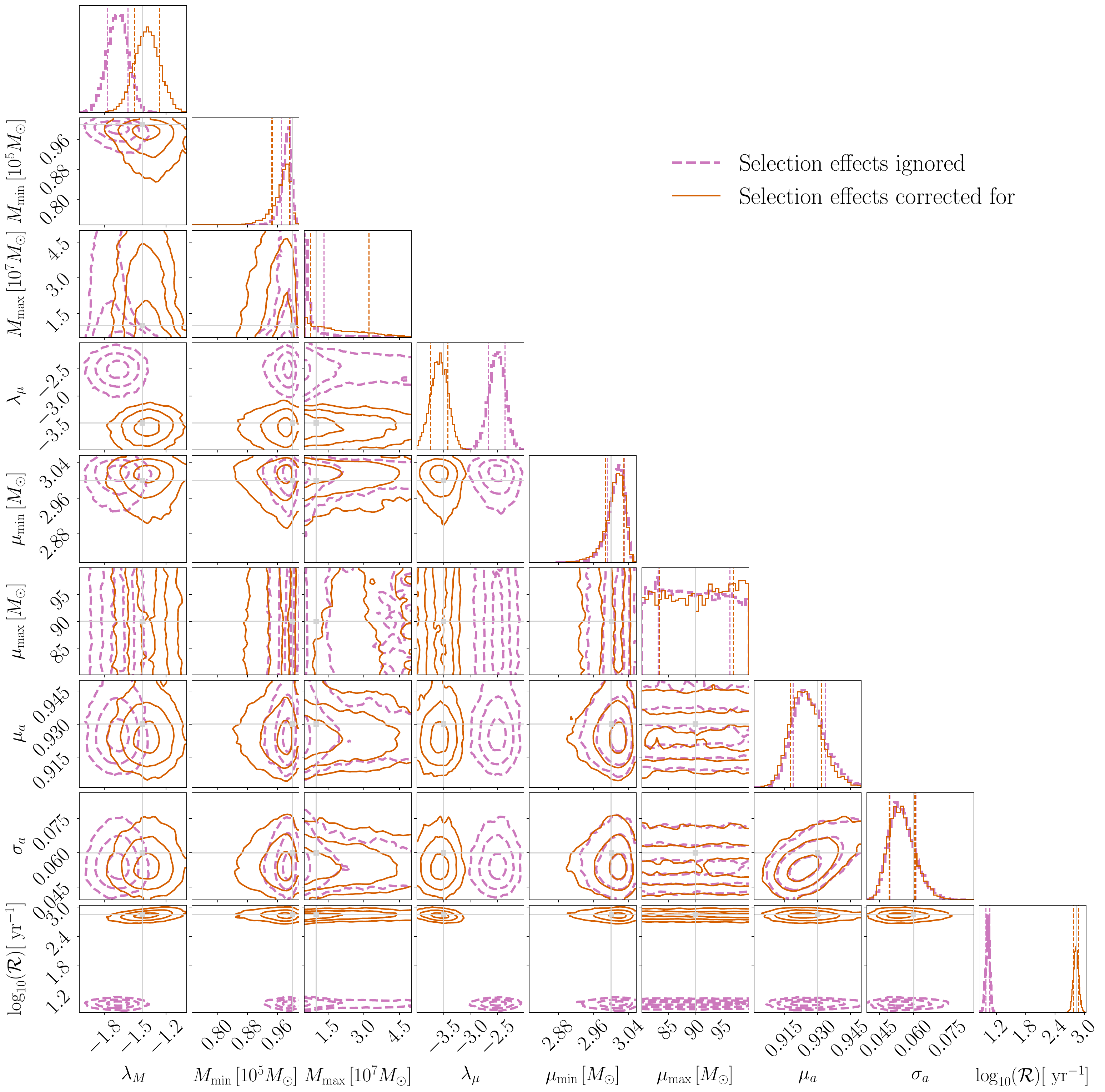}
    \caption{Full recovered hyperposteriors for our example scenario with selection effects corrected for (solid) or ignored (dashed). 
    The true values of each hyperparameter are indicated by the cross-hairs. 
    Significant bias is present for $\lambda_\mu$, with more minor bias for $\lambda_M$. 
    The rate $\mathcal{R}$ is accurately recovered with the inclusion of selection effect correction. 
    For other hyperparameters, the dashed hyperposteriors are over-constrained when compared to the corrected hyperposteriors.}
    \label{fig:joint_corner_full}
\end{figure*}

The full hyperposteriors from the population inference in Section~\ref{sec:sampling} are shown in Figure~\ref{fig:joint_corner_full}.
By including the mass range parameters, we can observe a more subtle consequence of the presence of selection biases: over-constrained hyperposteriors.
Neglecting selection effects leads to an underestimation of the error on parameters: this is reflected in the P--P plot analysis of Section~\ref{sec:pp}, where it is demonstrated that these effects lead to globally inconsistent hyperposterior effects at a statistically significant level.

\bsp
\label{lastpage}
\end{document}